\documentclass[5p,sort&compress,times,letterpaper]{elsarticle}
\usepackage{amsmath,amssymb}
\usepackage{graphicx}
\usepackage{hyperref}
\newcommand{\rmd}{\mathrm{d}}
\newcommand{\rmi}{\mathrm{i}}
\newcommand{\tr}{\mathrm{tr}}

\newcommand{\Gf}{G_\mathrm{F}}

\newcommand{\bnabla}{\boldsymbol{\nabla}}
\newcommand{\brho}{\protect{\bar\rho}}

\newcommand{\sfH}{\mathsf{H}} 
\newcommand{\bhv}{\boldsymbol{\hat{\mathbf{v}}}} 
\newcommand{\bfp}{\mathbf{p}} 
\newcommand{\bfx}{\mathbf{x}} 
 
\newcommand{\sfLambda}{\mathsf{\Lambda}} 
\newcommand{\sfD}{\mathsf{D}} 
\newcommand{\sfS}{\mathsf{S}}
\newcommand{\bfW}{\mathbf{W}}

\newcommand{\p}[1]{\protect{(#1)}}

\begin{document}

\title{Flavor instabilities in the neutrino line model} 

\author{Huaiyu Duan}
\ead{duan@unm.edu}
\author{Shashank Shalgar}
\ead{shashankshalgar@unm.edu}
\address{Department of Physics \& Astronomy, University of New
  Mexico, Albuquerque, NM 87131, USA}

\begin{abstract}
  A dense neutrino medium can experience 
  collective flavor oscillations
  through nonlinear neutrino-neutrino refraction.
  To make this multi-dimensional flavor transport problem more tractable,
  all existing studies have assumed certain symmetries
  (e.g., the spatial homogeneity and directional
  isotropy in the early universe) to reduce the dimensionality of the problem.
  In this work we show that, if both the
  directional and spatial
  symmetries are not enforced in the neutrino line
  model, collective oscillations can
  develop in the physical regimes where the symmetry-preserving
  oscillation modes
  are stable. Our results
  suggest that collective neutrino oscillations in real
  astrophysical environments (such as core-collapse supernovae and
  black-hole accretion discs) can be qualitatively different from the
  predictions based on existing models in which spatial and
  directional symmetries are artificially imposed. 
\end{abstract}

% \pacs{14.60.Pq, 97.60.Bw}

\maketitle

\section{Introduction} 
Neutrinos are influential in many hot and dense
astrophysical environments where they are copiously produced. For example,
$99\%$ of the immense power of a core-collapse supernova (SN) is carried
away by $\sim10^{58}$ neutrinos within just $\sim 10$ seconds (see,
e.g., \cite{Woosley:2005yv} for a review). Through the reactions
\begin{align}
  \nu_e + n &\rightleftharpoons p + e^-, &
  \bar\nu_e + p &\rightleftharpoons n + e^+
\end{align}
electron-flavor neutrinos extract energy from and 
deposit energy into the environment and change the $n$-to-$p$
ratio of the baryonic matter.

It has been firmly established by various experiments that neutrinos
can oscillate among different flavors or weak interaction states during
propagation. Most of the neutrino mixing
parameters have been determined, although it is still unknown whether
the neutrino has a normal mass hierarchy (NH) or an inverted one (IH),
i.e.\ whether 
$|\nu_3\rangle$ is the most massive of the three neutrino mass eigenstates
$|\nu_i\rangle$ ($i=1,2,3$) or not (see, e.g., 
\cite{Agashe:2014kda} for a review). The Mikheyev-Smirnov-Wolfenstein
(MSW) \cite{Wolfenstein:1977ue,Mikheyev:1985aa} 
flavor transformation of neutrinos in ordinary matter is also well
understood. However, because of its nonlinearity, our understanding of neutrino
oscillations in dense neutrino media (such as the one surrounding
the proto-neutron star in SN) is still elementary and requires more
investigation. 

In the absence of collisions the flavor evolution of neutrinos is
described by the Liouville equations
\cite{Sigl:1992fn,Strack:2005ux,Cardall:2007zw} 
\begin{subequations}
\label{eq:eom-full}
\begin{align}
\partial_t\rho + \bhv\cdot\bnabla\rho
&= -\rmi[\sfH_0 + \sfH_{\nu\nu},\, \rho],\\
\partial_t\brho + \bhv\cdot\bnabla\brho
&= -\rmi[\bar\sfH_0 + \sfH_{\nu\nu},\, \brho],
\end{align}
\end{subequations}
where $\bhv$ is the velocity of the neutrino, $\rho(t,\bfx,\bfp)$ is
the (Wigner-transformed) neutrino flavor density matrix which 
is a function of time 
$t$, position $\bfx$ and neutrino momentum $\bfp$, $\sfH_0$ is 
the Hamiltonian in the absence of ambient neutrinos,
and $\sfH_{\nu\nu}$ is 
the neutrino(-neutrino coupling) potential. Throughout this letter the
physical quantities with bars such as $\bar\rho$ and $\bar\sfH_0$ are
for antineutrinos. We assume that
neutrinos are relativistic so that $|\bhv|=1$ and neutrino energy
$E=|\bfp|$. The difficulty of solving Eq.~\eqref{eq:eom-full} 
stems from the neutrino potential which
couples neutrinos and antineutrinos of different momenta in
the neutrino medium:
\cite{Fuller:1987aa,Notzold:1987ik,Pantaleone:1992eq} 
\begin{align}
\sfH_{\nu\nu} = \sqrt2\Gf\int\frac{\rmd^3 p'}{(2\pi)^3}(1-\bhv\cdot\bhv')
[\rho(t,\bfx,\bfp')-\bar\rho(t,\bfx,\bfp')],
\label{eq:Hvv}
\end{align}
where $\Gf\approx(293\text{ GeV})^{-2}$ is the
Fermi coupling constant. When the neutrino potential is not
negligible, neutrinos in a dense medium can oscillate in a
collective manner (see, e.g.,
\cite{Duan:2010bg} for a review). In many cases collective
oscillations cause neutrinos of different flavors to swap their
spectra in certain energy ranges, a phenomenon dubbed as ``stepwise
spectral swap'' or ``spectral split'' (e.g.,
\cite{Duan:2006an,Duan:2006jv,Duan:2007sh,Dasgupta:2009mg}). 

Eq.~\eqref{eq:eom-full} poses a difficult 7-dimensional problem
with 1 temporal dimension, 3 spatial dimensions and 3 momentum dimensions.
All existing work on collective neutrino oscillations has assumed
certain directional 
symmetries in momentum space and/or spatial symmetries in position space to
make this problem more tractable. For example, the spatial spherical
symmetry and the directional axial symmetry (about the radial
direction) are generally assumed for SN
(e.g.,
\cite{Qian:1994wh,Pastor:2002we,Balantekin:2004ug,Duan:2006an,Duan:2006jv,Duan:2007sh,Dasgupta:2009mg,Duan:2010af,Mirizzi:2011tu,deGouvea:2012hg,Duan:2014mfa}),
and the spatial homogeneity
and directional isotropy for the early universe
(e.g., \cite{Kostelecky:1993yt, Abazajian:2002qx}).
However, these spatial and directional symmetries are not necessarily
preserved in collective neutrino oscillations. Imposing these
symmetries may lead to results that are qualitatively different from
those in real physical environments. It was recently
shown that collective oscillations can break the directional axial
symmetry in SN spontaneously \cite{Raffelt:2013rqa,Mirizzi:2013wda},
which obviously will not occur if this symmetry is artificially
enforced. Similar result is also found in the neutrino media
with an initial (approximate) isotropy \cite{Duan:2013kba}. A recent
numeric study shows that the spatial homogeneity can be also broken in
a toy model with 1 temporal and 1 spatial dimensions \cite{Mangano:2014zda}.

In this letter we propose to study the neutrino Line model
with 2 spatial dimensions. The results derived from this simple model
can provide valuable insights of the collective flavor transformation in
the neutrino gas models of multiple spatial dimensions.

\section{Neutrino Line model}
We consider the time-independent (neutrino) Line model in which
neutrinos are constantly emitted from 
the $x$ axis or the (neutrino) Line
and propagate inside the 
$x$-$z$ plane. For simplicity, we assume that every point on the Line
emits only 
neutrinos and antineutrinos of single energy $E$ with intensities $j_\nu$ and
$j_{\bar\nu}$, respectively, and in only two directions 
\begin{align}
  \bhv_\zeta = [u_\zeta,0,v_z] 
  \qquad(\zeta=L, R),
\end{align}
where $0<v_z<1$ and $u_R=-u_L=\sqrt{1-v_z^2}$. 
The Line model has 2 spatial dimensions $(x,z)$ and 2
momentum dimensions $(E,\zeta)$ (because an antineutrino of energy $E$ can be
treated as a neutrino of energy $-E$ for the purpose of 
neutrino oscillations).

We will consider the scenario of two flavor mixing, e.g., between
$\nu_e$ and $\nu_{\tau}$, in vacuum. In the mass basis
\begin{align}
  \sfH_0= -\bar\sfH_0 =
  -\frac{\omega\eta}{2}\sigma_3,
  \label{eq:H0}
\end{align}
where $\omega>0$ is the vacuum neutrino oscillation frequency,
$\eta=+1$ and $-1$ for NH and IH,
respectively, and
$\sigma_3$ is the third Pauli matrix.

We define reduced neutrino density matrices $\varrho\propto\rho$
and $\bar\varrho\propto\bar\rho$ which are normalized by condition
\begin{align}
  \tr\varrho = \tr\bar\varrho = 1.
\end{align}
The equations of motion for $\varrho$ and $\bar\varrho$ in the Line
model are
\begin{subequations}
\label{eq:eom}
\begin{align}
  \rmi \bhv_\zeta\cdot\bnabla\varrho_\zeta(x,z) &= [\sfH_0 +
    \sfH_{\nu\nu,\zeta}(x,z),\,    \varrho_\zeta(x,z)],\\
  \rmi \bhv_\zeta\cdot\bnabla\bar\varrho_\zeta(x,z) &= [-\sfH_0 +
    \sfH_{\nu\nu,\zeta}(x,z),\,    \bar\varrho_\zeta(x,z)].
\end{align}
\end{subequations}
The neutrino potential in the above equation is
\begin{align}
  \sfH_{\nu\nu,\zeta}(x,z) = \mu [\varrho_{\tilde\zeta}(x,z)
    - \alpha \bar\varrho_{\tilde\zeta}(x,z)],
\end{align}
where
\begin{align}
  \mu = \sqrt2 (1-\bhv_L\cdot\bhv_R) G_\text{F} j_\nu,
\end{align}
is the strength of the neutrino-neutrino coupling, 
$\tilde\zeta=R,L$ are the opposites of $\zeta$, and
$\alpha = j_{\bar\nu}/j_\nu$.

To facilitate numerical
calculations we further impose a periodic condition on the $x$-$z$
plane such that
\begin{align}
  \varrho_\zeta(x,z) &= \varrho_\zeta(x+L,z), &
  \bar\varrho_\zeta(x,z) &= \bar\varrho_\zeta(x+L,z),
\end{align}
where $L$ is the size of the periodic box. We define neutrino density
matrices in the Fourier space as
\begin{align}
  \varrho_{\zeta,m}(z) = \frac{1}{L}\int_0^L e^{-\rmi m k_0 x}
  \varrho_{\zeta}(x,z)\,\rmd x
  \label{eq:rho-m}
\end{align}
such that
\begin{align}
  \varrho_{\zeta}(x,z)&=\sum_m e^{\rmi m k_0 x} \varrho_{\zeta,m}(z),
\end{align}
where $m$ is an integer, and $k_0=2\pi/L$. 
We also define $\bar\varrho_{\zeta,m}(z)$ for the antineutrino in a
similar way. Using Eq.~\eqref{eq:eom} and
\begin{align}
  \bhv_\zeta\cdot\bnabla\varrho_\zeta &= \sum_m e^{\rmi m k_0 x}
      [v_z\varrho'_{\zeta,m} 
        +\rmi m k_0 u_\zeta \varrho_{\zeta,m}]
\end{align}
we obtain the equations of motion in the Fourier space:
\begin{subequations}
\label{eq:eom-m}
\begin{align}
  \rmi v_z \varrho'_{\zeta,m} &=
  m k_0 u_\zeta \varrho_{\zeta,m}
  + [\eta\omega\sigma_3/2,\, \varrho_{\zeta,m}]
  \nonumber\\
&\quad  + \mu \sum_{m'}[ \varrho_{\tilde\zeta,m'}-\alpha
  \bar\varrho_{\tilde\zeta,m'},\, \varrho_{\zeta,m-m'}],\\
  \rmi v_z \bar\varrho'_{\zeta,m} &=
  m k_0 u_\zeta \bar\varrho_{\zeta,m}
  + [-\eta\omega\sigma_3/2,\, \bar\varrho_{\zeta,m}]
  \nonumber\\
& \quad + \mu \sum_{m'}[ \varrho_{\tilde\zeta,m'}-\alpha
  \bar\varrho_{\tilde\zeta,m'},\, \bar\varrho_{\zeta,m-m'}],
\end{align}
\end{subequations}
where $\varrho'_{\zeta,m}=\rmd\varrho_{\zeta,m}/\rmd z$.

\section{Flavor instabilities}
It is instructive to first review the flavor instability in the
bipolar model with 1 spatial (or
temporal) dimension and 1 momentum dimension
\cite{Kostelecky:1994dt,Duan:2005cp,Hannestad:2006nj,Duan:2007mv}.  
This model can be obtained from the Line model by imposing
the translation symmetry along the $x$ axis and the
left-right symmetry ($L\leftrightarrow R$) between the two angle directions.
For the bipolar model Eq.~\eqref{eq:eom} has solution
\begin{subequations}
\begin{align}
  \varrho_{\zeta}(x,z) &=
  e^{-\rmi \omega z \eta\sigma_3/2 v_z}
  \varrho(0)
  e^{\rmi \omega z \eta\sigma_3/2 v_z}, \\
  \bar\varrho_{\zeta}(x,z) &=
  e^{-\rmi (-\omega) z \eta\sigma_3/2 v_z}
  \bar\varrho(0)
  e^{\rmi (-\omega) z \eta\sigma_3/2 v_z}
\end{align}
\end{subequations}
in the absence of ambient neutrinos (i.e.\ $\mu=0$),
where $\varrho(0)$ and $\bar\varrho(0)$ are the neutrino density
matrices at $z=0$.
In this vacuum oscillation solution an antineutrino behaves as a
neutrino with a negative oscillation frequency $-\omega$ or negative
energy $-E$. Inside the
neutrino medium, however, $\varrho$ and
$\bar\varrho$ can oscillate with the same frequency $\Omega$ 
under suitable conditions such that 
\begin{subequations}
\label{eq:coll-sol}
\begin{align}
  \varrho_{\zeta}(x,z) &=
  e^{-\rmi \Omega z \eta\sigma_3/2 v_z}
  \varrho(0)
  e^{\rmi \Omega z \eta\sigma_3/2 v_z}, \\
  \bar\varrho_{\zeta}(x,z) &=
  e^{-\rmi \Omega z \eta\sigma_3/2 v_z}
  \bar\varrho(0)
  e^{\rmi \Omega z \eta\sigma_3/2 v_z},
\end{align}
\end{subequations}
where $\Omega$ is a function of $\mu$, $\alpha$ and $\omega$
\cite{Duan:2007mv}. This solution is equivalent to the precession motion
of a pendulum in flavor space \cite{Hannestad:2006nj}.
Similar solutions can exist for the scenarios with a
continuous energy distribution of neutrinos \cite{Raffelt:2007cb}.
When the neutrino mixing angle $\theta$ is small, the
collective oscillation 
solution in Eq.~\eqref{eq:coll-sol} does not result in significant
neutrino oscillations unless $\kappa=\text{Im}(\Omega)>0$. A positive
$\kappa$ indicates that the flavor pendulum cannot precess stably and must
experience nutation in flavor space. This flavor instability can lead to
collective neutrino oscillations with observable effects.

For the Line model the collective solution should take the form
\begin{subequations}
\begin{align}
  \varrho_{\zeta}(x,z) &=
  e^{-\rmi \Omega z \eta\sigma_3/2 v_z}
  \varrho(x_0,0)
  e^{\rmi \Omega z \eta\sigma_3/2 v_z}, \\
  \bar\varrho_{\zeta}(x,z) &=
  e^{-\rmi \Omega z \eta\sigma_3/2 v_z}
  \bar\varrho(x_0,0)
  e^{\rmi \Omega z \eta\sigma_3/2 v_z},
\end{align}
\end{subequations}
where $(x_0=x-u_\zeta z/v_z, z_0=0)$ is the coordinate of the emission
point of the neutrino  
which propagates in direction $\bhv_\zeta$ and passes through $(x,z)$.
To study the flavor stability in the Line model
we assume that the neutrino mixing angle $\theta\ll1$
and that neutrinos and antineutrinos are in almost pure
electron flavor:
\begin{align}
  \varrho &\approx \begin{bmatrix}
    1 & \epsilon \\ \epsilon^* & 0
    \end{bmatrix}, &
    \bar\varrho &\approx \begin{bmatrix}
    1 & \bar\epsilon \\ \bar\epsilon^* & 0
    \end{bmatrix},
\end{align}
where $|\epsilon|\sim|\bar\epsilon|\ll1$.
The flavor instability of the
neutrino medium can be 
studied by using the method of linearized stability analysis
\cite{Banerjee:2011fj}.  
For this purpose we define
\begin{subequations}
\begin{align}
  \sfD_m^\pm &= \frac{1}{2}(\varrho_{L,m}\pm\varrho_{R,m})
  -\frac{\alpha}{2}(\bar\varrho_{L,m}\pm\bar\varrho_{R,m})
  \nonumber\\
  &\approx \begin{bmatrix}
    (1 - \alpha)\,\delta_{m,0} & D^\pm_m \\ (D^\pm_{-m})^* & 0
  \end{bmatrix},\\  
  \sfS_m^\pm &= \frac{1}{2}(\varrho_{L,m}\pm\varrho_{R,m})
  +\frac{\alpha}{2}(\bar\varrho_{L,m}\pm\bar\varrho_{R,m})
  \nonumber\\
  &\approx \begin{bmatrix}
    (1 + \alpha)\,\delta_{m,0} & S^\pm_m \\ (S^\pm_{-m})^* & 0
  \end{bmatrix}.
\end{align}
\end{subequations}
Keeping only the terms up to
$\mathcal{O}(|\epsilon|)$ in Eq.~\eqref{eq:eom-m} we obtain
\begin{align}
  \rmi\frac{\rmd}{\rmd z}\bfW_m(z) = \sfLambda_m\cdot\bfW_m(z),
  \label{eq:eom-W}
\end{align}
where $\bfW_m(z) = [D_m^+,\, S_m^+,\,  D_m^-,\, S_m^-]^T$, and
\begin{align}
  \sfLambda_m = v_z^{-1}\begin{bmatrix}
    0 & -\eta\omega & m q& 0 \\
    -\eta\omega-\mu_+ & \mu_- & 0 & m q\\
    m q& 0 & 2 \mu_- & -\eta\omega \\
    0 & m q& -\eta\omega + \mu_+ & \mu_-
  \end{bmatrix}
  \label{eq:Lambda}
\end{align}
with $q=(2\pi/L)\sqrt{1-v_z^2}$ and $\mu_\pm = (1\pm\alpha)\mu$.

Eq.~\eqref{eq:eom-W} shows that in the Line model different Fourier
modes are decoupled in the linear regime.
In general the real matrix $\sfLambda_m$ has four eigenvalues
$\Omega_m^\p{i}$ ($i=1,2,3,4$). If $\Omega_m^\p{i}$ is complex, then its
complex conjugate is also an eigenvalue of $\sfLambda_m$. 
Because the characteristic polynomial of
$\sfLambda_m$ contains only even powers of $m$, the eigenvalues of
$\sfLambda_{-m}$ are the same as those of $\sfLambda_m$. Further,
the set of values of $\kappa_m^\p{i}=\text{Im}[\Omega_m^\p{i}]$ for
given $m$ and $\mu$ is 
independent of the neutrino mass hierarchy in the Line model because
$\sfLambda_m\rightarrow 2(\mu_-/v_z)\mathsf{I} - \sfLambda_{-m}$
under transformation
\begin{align*}
  \eta &\rightarrow -\eta, &
  (D_m^\pm,S_m^\pm) &\rightarrow (D_m^\mp,S_m^\mp).
\end{align*}

\begin{figure*}[ht]
  \begin{center}
    $\begin{array}{@{}c@{\hspace{0.1in}}c@{}}
      \includegraphics*[scale=0.78]{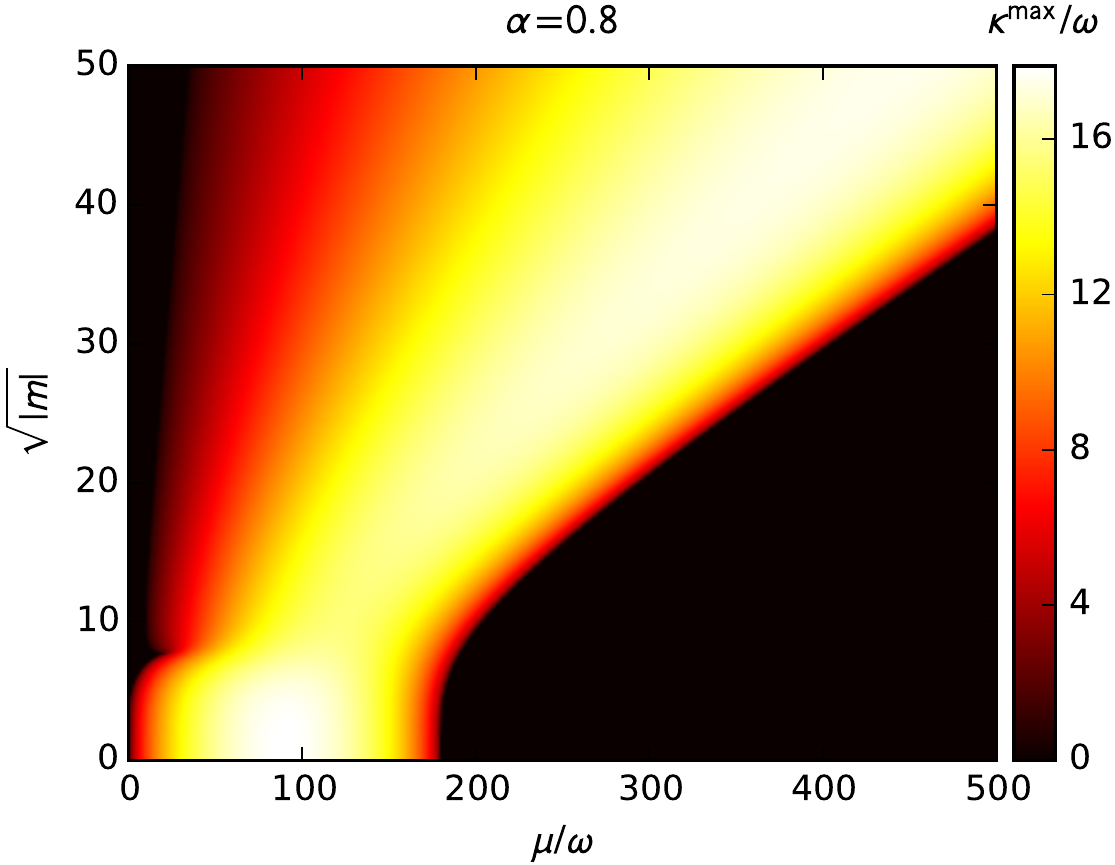} &
      \includegraphics*[scale=0.78]{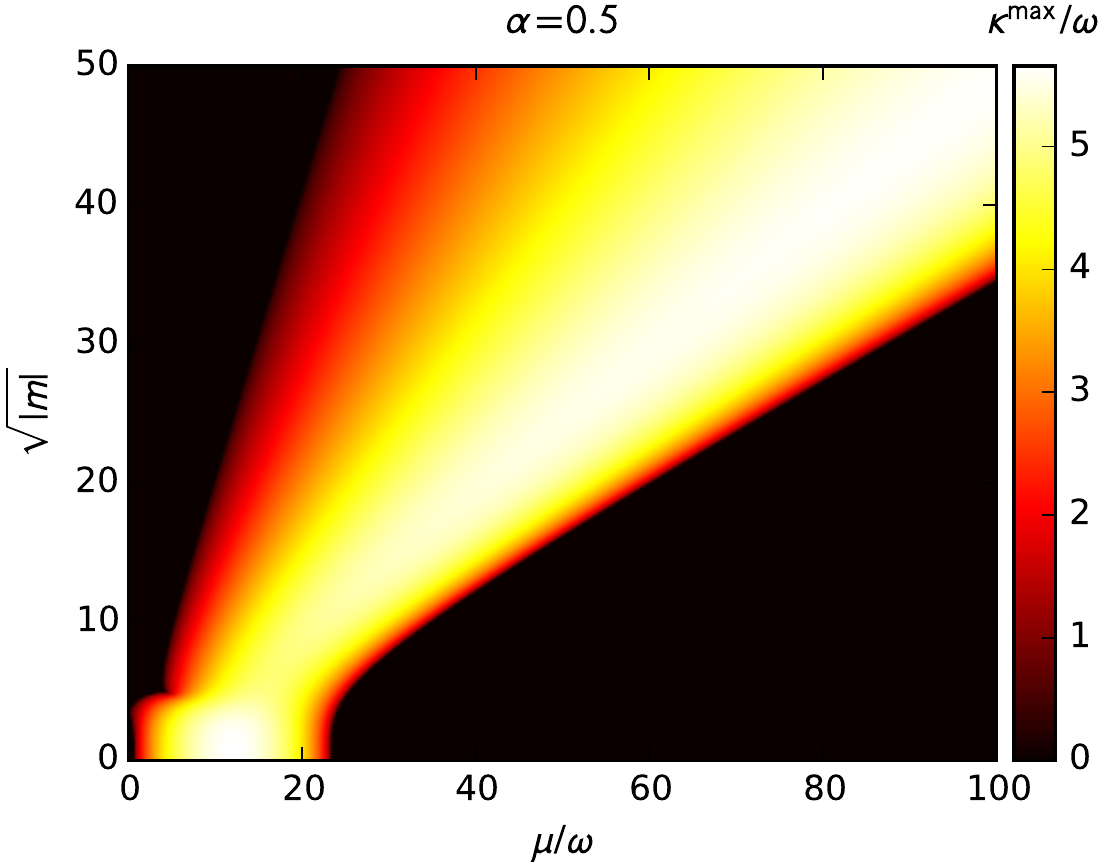}
      \end{array}$
  \end{center}
  \caption{The flavor stability of the two-dimensional ($x$-$z$),
    two-angle, mono-energetic neutrino gas in the parameter space of neutrino
    self-coupling strength $\mu$, which is proportional to the
    neutrino number density, and moment index $m$.
    The color scale of the plots
    represents $\kappa_m^\text{max}(\mu)$, the largest exponential
    growth rate of the corresponding collective mode of neutrino
    oscillations in terms of $z$ for given $\mu$ and $m$.
    Both $\mu$ and $\kappa_m^\text{max}(\mu)$ are measured in terms of
    the oscillation frequency $\omega$ of the neutrino in vacuum.
    Collective
    oscillation modes with $m=0$ preserve the translation symmetry
    along the $x$ direction, but those with $m\neq0$ break this
    symmetry spontaneously. The larger the value of $|m|$, the smaller
    scales are the  
    flavor structures in neutrino fluxes.
    The ratio of the number flux of antineutrinos to that of
    neutrinos is $\alpha=0.8$ in the left panel and $0.5$ in the right
    panel. In both panels,
    the size of the periodic box of the neutrino sources on the
    $x$-axis is $L=20\pi/\omega$, and the propagation directions of the
    neutrinos are given by unit vectors $[v_x,v_z]=[\pm\sqrt3/2,1/2]$
    which make $60^\circ$ angle with the $z$ axis. The
    results are independent of the neutrino mass hierarchy.}
\label{fig:kappa}
\end{figure*}

In Fig.~\ref{fig:kappa} we plot
$\kappa_m^\text{max}(\mu)$, the largest of all $\kappa_m^\p{i}(\mu)$, as
functions of $\mu$ and $m$ for the cases with
$v_z=0.5$, $L=20\pi/\omega$ and $\alpha=0.8$ and $0.5$, respectively.
Fig.~\ref{fig:kappa} shows that collective oscillation modes of 
different $|m|$ values are unstable in
different physical regimes in the Line
model.
As $|m|$ increases, the flavor unstable region shifts to larger $\mu$,
and its width also increases.
In addition,
as the asymmetry in the number fluxes of neutrinos and antineutrinos
decreases, the flavor unstable regions move to larger $\mu$, and the
collective oscillation modes develop faster (because
$\kappa_m^\text{max}$ are larger).

\section{Discussion}

In the Line model we have assumed two initial approximate symmetries:
the (spatial) translation symmetry 
along the $x$ axis in position space and the
(directional) left-right symmetry 
between the two neutrino beams in momentum space.
As mentioned previously, when both symmetries are
strictly imposed, the Line model reduces to the bipolar model.
Albeit a simple model, the bipolar model has shed important insights
on the early simulations of supernova neutrino 
oscillations in the (neutrino) Bulb model \cite{Hannestad:2006nj,Duan:2007mv},
which seems perplexing at the first sight \cite{Duan:2006an,Duan:2006jv}.
For example, because only the homogeneous and
(left-right) symmetric mode 
$[D_0^+,S_0^+]$ can exist in the bipolar model, 
Eq.~\eqref{eq:Lambda} shows that collective oscillations can develop
only in IH and in the regime
\begin{align}
  \frac{2\omega}{(1+\sqrt\alpha)^2} < \mu <
  \frac{2\omega}{(1-\sqrt\alpha)^2}.
  \label{eq:range}
\end{align}
Indeed, unless the MSW transformation has significantly changed
neutrino energy spectra,
collective neutrino oscillations in the Bulb model with
bipolar-like neutrino fluxes (i.e.\ dominated by the $\nu_e$ and
$\bar\nu_e$) occur only in IH and within certain radius range
\cite{Duan:2006an}.
We note that, like the bipolar model, the Bulb model also has
a symmetry in position space and one in momentum space: the
spatial spherical symmetry around the center of the proto-neutron star and the
directional axial symmetry about the radial direction. This similarity is the
reason why the bipolar and Bulb models can produce qualitatively
similar results although their geometries are quite different.

The two-beam model proposed in  \cite{Raffelt:2013isa} is one step
away from the bipolar model where the 
left-right symmetry in momentum space is not enforced.
Eq.~\eqref{eq:Lambda} shows that,
because of the availability of the anti-symmetric mode
$[D_0^-,S_0^-]$, collective oscillations can occur also in NH and in the same
regime described by Eq.~\eqref{eq:range}.
The two-beam model was helpful in
understanding collective neutrino oscillations in the extended Bulb model
in which the axial symmetry in momentum space is not imposed and the
axial-symmetry-breaking modes become unstable in NH for bipolar-like
neutrino fluxes
\cite{Raffelt:2013rqa,Mirizzi:2013rla}. 
Again, the two-beam model and the extend Bulb model can produce
qualitatively similar results because a directional symmetry in
momentum space is broken in both models.

The toy model studied in \cite{Mangano:2014zda} is equivalent to the
Line model with the left-right symmetry and $\alpha=1$. In this toy
model only the symmetric modes are available.
Eq.~\eqref{eq:Lambda} shows that the inhomogeneous (i.e.\ with $m\neq0$)
symmetric modes $[D_{m\neq0}^+, S_{m\neq0}^+]$ in this model are
unstable in the same regime as the 
homogeneous symmetric mode does in the bipolar model. Indeed,
the numerical calculations in \cite{Mangano:2014zda} show that a
slight matter inhomogeneity can lead to collective oscillations that
break the spatial symmetry.

In the Line model neither the spatial symmetry or the directional
symmetry is imposed. Because the inhomogeneous symmetric and anti-symmetric
modes are coupled even in the linear regime, both the
spatial and directional symmetries are generally broken in the Line
model, and collective oscillations can occur in both NH and IH.
What is more intriguing is that
the inhomogeneous modes can be unstable in the regimes
of higher neutrino fluxes than the homogeneous modes.

As in the case of bipolar and
two-beam models, the results obtained in the Line model may provide
valuable insights of collective neutrino oscillations in multi-dimensional
models for core-collapse supernovae or black-hole accretion discs
(e.g., \cite{Malkus:2012ts,Malkus:2014iqa}). It seems very likely that
the spherical symmetry 
about the center of the supernova or the axial symmetry about the
central axis of the black-hole accretion disc can be broken by
collective neutrino oscillations even if such spatial symmetries hold
approximately at first. 
The results of the Line model also suggest
that collective neutrino oscillations can 
occur in astrophysical environments in the regions of higher 
neutrino fluxes than what is predicted by symmetry-preserving models.
This can lead a larger impact on nucleosynthesis in these
environments than previously expected (e.g.,
\cite{Pastor:2002we,Balantekin:2004ug,Chakraborty:2009ej,Duan:2010af}).  

In the Line model it seems that the central value $\mu$ of the flavor unstable
region increases linearly with $\sqrt{|m|}$ when $|m|$ is sufficiently
large. In a real physical system there must 
exists a cutoff value $m_\text{max}$ so that the inhomogeneous modes with
$|m|>m_\text{max}$ are suppressed. After all, 
Eq.~\eqref{eq:eom} becomes invalid for too small length scales because it is
based on the assumption of coherent forward scattering of neutrinos by
the medium. 

Although we have assumed zero matter density in our discussion, the
results also apply to the scenario with large matter density. In the
latter case the co-rotating frame technique can be used to ``remove''
the effects of the matter density on collective neutrino oscillations
\cite{Duan:2005cp}. 

\section{Conclusions}
We have shown that collective neutrino oscillations can break both the
spatial and directional symmetries in the neutrino Line
model. We found that inhomogeneous neutrino oscillation modes can
become unstable in the regimes where the homogeneous modes are stable.
Our results suggest that collective neutrino oscillations in real
astrophysical environments can be qualitatively different from the
predictions based on the models with artificially imposed spatial and
directional symmetries. 

Our results also suggest that collective oscillations in a
multi-dimensional neutrino gas model can be highly inhomogeneous. 
The large inhomogeneity can pose a great challenge to such
studies in both computing resources and numerical modeling.

The Line model which we have studied here has only two neutrino beams from
each emission point. It will be interesting to see whether the
inhomogeneous modes can 
be suppressed in the regions of very high matter or neutrino density
due to the multi-angle suppression when a 
continuous angle distribution of neutrino fluxes are employed 
\cite{EstebanPretel:2008ni,Duan:2010bf}. 

\section*{Acknowledgments}
  We thank S.~Abbar and S.~V.~Noormofidi for useful
  discussions. We also thank A.~Mirizzi for bringing
  Ref.~\cite{Mangano:2014zda} to our attention.
  This work was supported by DOE EPSCoR grant \#DE-SC0008142 at UNM.

\section*{References}
\bibliographystyle{elsarticle-num}
\bibliography{line2a}

\end{document}